\documentclass[submit]{aiaa-tc}

\usepackage{amsmath,graphicx,amssymb}

 \usepackage{varioref}
 \usepackage{wrapfig}
 \usepackage{threeparttable}
 \usepackage{dcolumn}
  \newcolumntype{d}{D{.}{.}{-1}}
 \usepackage{nomencl}
  \makeglossary
 \usepackage{subfigure}
 \usepackage{subfigmat}
 \usepackage{fancyvrb}
  \fvset{fontsize=\footnotesize,xleftmargin=2em}
 \usepackage{lettrine}
 \usepackage[colorlinks]{hyperref}

 \title{Towards a priori uncertainty quantification in coarse-grained molecular dynamics: \\ Generalized multipole potentials}

 \author{
  Paul N. Patrone  \thanks{Applied and Computational Mathematics Division, National Institute of Standards and Technology, 100 Bureau Drive, Gaithersburg MD, 20899, USA.}, Andrew M. Dienstfrey \thanksibid{1},
  Geoffrey B. McFadden  \thanksibid{1}
 }

\def\br{\boldsymbol {\rm r}}

\def\bx{\boldsymbol {\rm \chi}}

\def\bt{\boldsymbol \Theta}
\def\boldx{\boldsymbol {\rm x}}

\def\T{{\rm T}}

\def\blk{\boldsymbol{ \delta}_k}
\def\blkp{\boldsymbol\delta_{k'}}

\def\Tr{{\rm Tr}}

\def\U{\mathcal U}
\def\bomega{\boldsymbol \omega}

\def\bchi{\boldsymbol \chi}
\def\bDelta{\boldsymbol \Delta}
\def\C{\boldsymbol { \mathcal C}}

\def\K{\boldsymbol {\mathcal K}}

\begin{document}

\maketitle

\begin{abstract}
In computational materials science, coarse-graining approaches often lack a priori uncertainty quantification (UQ) tools that estimate the accuracy of a reduced-order model before it is  calibrated or deployed.  This is especially the case in coarse-grained (CG) molecular dynamics (MD), where ``bottom-up'' methods need to run expensive atomistic simulations as part of the calibration process.  As a result, scientists have been slow to adopt CG techniques in many settings because they do not know in advance whether the cost of developing the CG model is justified.  To address this problem, we present an analytical method of coarse-graining rigid-body systems that yields corresponding intermolecular potentials with controllable levels of accuracy relative to their atomistic counterparts.  Critically, this analysis: (i) provides a mathematical foundation for assessing the quality of a CG force field without running simulations; and (ii) provides a tool for understanding how atomistic systems can be viewed as appropriate limits of reduced-order models.  Simulated results confirm the validity of this approach at the trajectory level and point to issues that must be addressed in coarse-graining fully non-rigid systems.  
\end{abstract}


\lettrine[nindent=0pt]{S}{cientists} have been slow to adopt  so-called ``bottom-up'' coarse-grained (CG) molecular dynamics (MD) in  materials development and integrated computational materials engineering (ICME) settings.  From a practical perspective, this can be traced to the fact that most, if not all, CG-MD strategies lack {\it a priori} uncertainty quantification (UQ) tools that estimate the accuracy of a CG model relative to its atomistic counterpart \cite{transfer1,represent1}.  As a result, modelers do not know beforehand whether the cost to develop a CG model justifies the end result and have therefore been reluctant to deploy such tools without outside validation.  Just as troubling, a variety of studies meant to provide exactly this validation have instead shown that many CG methods  suffer from an inability to correctly predict more than one or two material properties, and even then, only after significant recalibration \cite{PatroneRosch16,represent1,transfer1,transfer2,transfer3,transfer4,rentropy}.  Seen collectively, these issues paint a confusing picture for the viability of CG-MD and suggest the need for a deeper mathematical understanding of how these methods work. 

Conceptually, the lack of {\it a priori} UQ can be understood through comparison with coarse-graining strategies used in other fields, e.g.\ electrostatics.  Multipole expansions provide an especially compelling example.\cite{jackson12}  In this case, complicated charge distributions are treated as a weighted sum over much simpler geometric configurations (i.e.\ dipoles, quadrupoles, octopoles, etc.) under far-field conditions.  The three main benefits of this approach are that: (i) multipole interactions are  pre-computed, so that there is no need to evaluate complicated Coulomb interaction integrals; (ii) truncations (i.e.\ ``coarse-grained'' representations) of the multipole expansion have errors that are relatively straightforward to bound; and (iii) the associated expansion converges to the full interaction in as we keep successively more terms.  These last two properties are especially important because they imply an analytical connection between the coarse-grained and full model that can be exploited to both quantify and control limitations of the former.  

By contrast, many CG-MD strategies tend to rely on {\it ad hoc} coordinate reduction strategies that lack a notion of limiting procedure or related connection to the atomistic model.\cite{rentropy,Reith2003,mscg07,review1}  Rather, the key idea is to: (i) define a projection operator that maps the detailed atomistic system onto the desired CG representation; and then (ii) numerically compute a corresponding potential by optimizing an objective that compares the atomistic and CG predictions.  As a result, the atomistic system cannot be shown to arise as a limiting case of the CG model, since the latter is {\it chosen} by the user and not defined through perturbative or limiting-type arguments.  Moreover, notions of trajectory-level convergence between models is lost, since the projection is only enforced in an average sense.  As a result, such methods had defied attempts to provide rigorous and {\it a priori} UQ to date, especially in the realm of non-equilibrium statistical mechanics.\cite{represent1}  

As a first step to addressing these problems, we present a coarse-graining strategy that reformulates atomistic potentials of rigid bodies in terms of generalized multipole expansions and treats CG models as truncations thereof.    The main intuition underlying this analysis is the observation that a rigid body is characterized by only six degrees of freedom, namely its center-of-mass (COM) position and orientation.  As a result, the corresponding intermolecular potential collapses to a sum of geometric moments of the rigid body that interact in an orientation-dependent manner. This has the benefit of allowing us to quantify errors in a CG potential without ever having to run a simulation by analyzing truncations of the atomistic potential.  Moreover, it provides an {\it analytical} route to understanding the limitations of other CG-MD approaches that invoke alternate coordinate reduction strategies.  Simulated results at the trajectory level confirm our main results and point to open problems.

We emphasize that our approach, while analytical in nature, has certain limitations.  First, accuracy of the method relies on a far-field approximation, which is obviously not valid for nearby particles.  Second, our method, as developed, only applies to rigid molecules, which is not a good approximation for all systems.  In the discussion section, we pursue these issues in more detail and point towards the appropriate mathematical resolutions.  

The rest of this paper is organized as follows.  In the next section, we present our generalized multipole potentials for simple power-law systems.  The following section presents simulated results, with a discussion and conclusion following thereafter.  

\section{Generalized Multipole Potentials}

For simplicity, we consider a system composed of molecules having $P$ identical sub-components.\footnote{Here we mean identical in the sense that all particles have the same interaction law, but not necessarily the same mass.}  Generalizations to more realistic systems are straightforward but require burdensome notation that is not helpful in an expository setting.  To fix notation, we posit that the atomic positions in a molecule are given by 
\begin{align}
\boldx_k  = \bx + \bt \blk, \label{eq:cgrep}
\end{align}
where $\boldx_k$ is the atomic position in a laboratory frame of reference, $\bx$ is the corresponding center of mass coordinate, $\blk$ is the coordinate of the atom in a fixed body reference frame, and $\bt$ is a rotation matrix connecting the body and lab frames.  It is important to note that Eq.~\eqref{eq:cgrep} is a coarse-grained representation of a molecule in the sense that $\blk$ is fixed, so that the atomistic coordinates are specified entirely in terms of $\bx$ and $\bt$.  Moreover, $\bt$ only has three independent degrees-of-freedom (DOF), so that each CG molecule has a total of six independent DOF.   

To make use of this decomposition, we consider as an example the case in which the interatomic interaction is given by $u(r) = r^{-n}$, where $r$ is the separation distance between two atoms on different molecules.  The intermolecular interaction between molecules $j$ and $j'$ is then given by
\begin{align}
\U &= \sum_{k,k'=1}^P \frac{1}{[\chi^2 + 2 \bx \cdot \bDelta_{k,k'} + \Delta_{k,k'}^2]^{n/2}} \label{eq:inversepower},
\end{align}
where $k$ ($k'$) indexes molecules in molecule $j$ ($j'$), $\bx=\bx_j - \bx_{j'}$, $\chi = |\bchi|$ and $\bDelta_{k,k'} =\bt_j\blk - \bt_{j'}\blkp$, and $\Delta_{k,k'}=|\bDelta_{k,k'}|$.  In order to make the following expressions more compact, we define $\phi_1 = 2 \bx \cdot \bDelta_{k,k'} / \chi$ and $\phi_2 = \Delta_{k,k'}^2$.    By analogy with multipole expansions in electrostatics, Eq.~\eqref{eq:inversepower} can then be expressed as a series expansion of the form
\begin{align}
\U &= \sum_{k,k'} \frac{1}{\chi^n}\left[1+ \frac{\phi_1}{\chi} + \frac{\phi_2}{\chi^2}  \right]^{-n/2} \nonumber \\
&=\frac{P^2}{\chi^n} +\frac{1}{\chi^n}\sum_{k,k'} \sum_{q=1}^{\infty} \frac{\Gamma\left(\frac{n}{2} + q  \right)}{\Gamma\left(\frac{n}{2} \right) q!} \left[\frac{-\phi_1}{\chi} - \frac{\phi_2}{\chi^2} \right]^q \label{eq:expansion1}
\end{align}
where we assume that $|\phi_1/\chi + \phi_2/\chi^2| < 1$, i.e.\ the intermolecular separation is larger than the intramolecular distances.  Note that $\Gamma(x)$ is the gamma function. Interchanging sums over $k,k'$ and $q$, we find that the potential can be expressed in the form
\begin{align}
\U = \frac{P^2}{\chi^n} + \frac{1}{\chi^n}\sum_{q} \frac{U_q(\bx,\bt_j,\bt_{j'})}{\chi^q} \label{eq:expansion2}
\end{align}
where the $U_q$ no longer require $P^2$ evaluations over the atomic coordinates.\cite{Patroneinprep}

To see this last point in more detail, consider the $\mathcal O(\chi^{-n-1})$ term in Eq.~\eqref{eq:expansion2}.  One finds that 
\begin{align}
U_1 & = -\frac{n}{2} \sum_{k,k'} \phi_1 \nonumber \\
&= -n \frac{\bx^\T}{\chi} \left[\bt_j \sum_{k,k'} \blk - \bt_{j'}\sum_{k,k'} \blkp \right] \nonumber \\
&= -nP \frac{\bx^\T}{\chi} [\bt_j - \bt_{j'}] \C \label{eq:foc}
\end{align}
where $\C$ is the centroid defined by
\begin{align}
\C = \sum_k \blk.
\end{align}
Note that $\C$ can be precomputed for all molecules, so that Eq.~\eqref{eq:foc} is an inner product weighted by a difference of rotation matrices.  The second order term is given by 
\begin{align}
U_2 &= \sum_{k,k'} \frac{-n}{2}\phi_2 + \left(\frac{n}{2} + 1 \right) \frac{n}{4}\phi_1^2 \nonumber \\
&=-n \left[P \, \Tr(\K)  - \C^\T \bt_j^\T \bt_{j'} \C\right],\nonumber \\
& \,\,\,\,+ \left(\frac{n}{2} + 1 \right) n \frac{\bx^\T}{\chi} \Big[P\bt_j \K \bt_j^\T + P\bt_{j'} \K \bt_{j'}^\T \nonumber \\ 
&\qquad \qquad\qquad \qquad \qquad-  2 \bt_j \C \C^\T \bt_{j'}^\T\Big]  \frac{\bx}{\chi} \label{eq:soc}
\end{align}
where $\K$ is a sum of tensor products given by
\begin{align}
\K = \sum_{k}\blk \blk^{\rm T}
\end{align}
if $\blk$ is expressed as a column vector.  Higher order corrections are more tedious to write explicitly and involve higher-order tensor products of the body-frame coordinates.\cite{Patroneinprep}  

It is important to note that in using Eqs.~\eqref{eq:expansion2}--\eqref{eq:soc} in MD simulations, one requires explicit expressions for the force and torque, which likewise depend on $\bx$ and $\bt$.  Generalized formulas for these expressions have been computed in many works, to which we refer to reader for more details.\cite{Patroneinprep,Limecooler,Limecooler2}  We also note that this analysis works for any smooth potential, not just power laws.  In particular, if $u(r)$ is an arbitrary but differentiable interatomic potential, the corresponding generalized multipole potential is given by
\begin{align}
\U = u(\bx) + \sum_q \sum_{k,k'} u^{(q)}(\bx) \frac{(2\bx \cdot \bDelta_{k,k'} + \bDelta_{k,k'}^2)^q}{q!},
\end{align}
where $u^{(q)}$ denotes the $q$th derivative of $u$ with respect to $\bx$.  Simplifying the sums over $k,k'$ yields the corresponding multipole expansions.

\section{Simulated Results}

\begin{figure*}[ht]
\includegraphics[width=18cm]{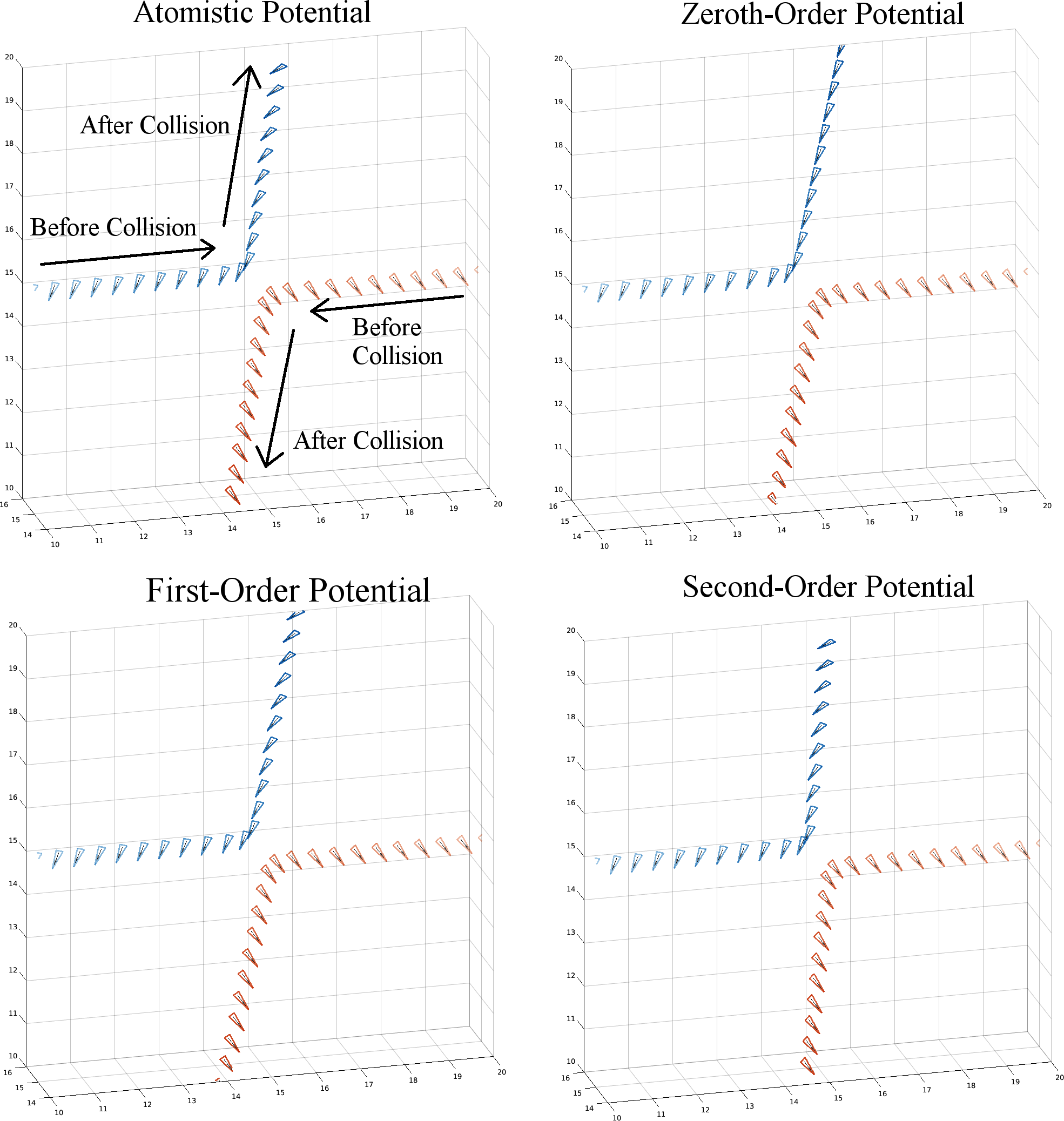}\caption{Example of predictions associated with an atomistic intermolecular potential and approximations thereof.  The four subplots show collisions between two triangular particles.  {\it Top left:} Trajectory computed using the atomistic potential.  Note that the triangular molecules begin rotating after the collision.  {\it Top right:}  Trajectory using the zeroth order (radially symmetric) potential.  Note that the collision does not induce any rotation in the molecules.  {\it Bottom left:}  Trajectory using the first order potential.  {\it Bottom right:}  Trajectory computed using the second order potential.}\label{fig:collsion}
\end{figure*}

To illustrate some of the properties of the multipole potentials, we consider an idealized system composed of triangular rigid bodies with interaction sites located at the verticies.  We consider isosceles molecules that have different masses at each vertex, so that the center-of-mass does not coincide with the centroid.  In the body-frame, these verticies are located at
\begin{subequations}
\begin{align}
\boldsymbol \delta_1 &= ab(1,0,0)^\T \label{eq:tr1} \\
\boldsymbol \delta_2 &= b(-0.5 ,\sqrt{3}/2,0)^\T \\
\boldsymbol \delta_3 &= b(-0.5,-\sqrt{3}/2,0)^\T \label{eq:tr2} 
\end{align}  
\end{subequations}
where $a$ determines the relative length of the two equal sides to the remaining side of the triangle, and $b$ is an overall scale factor determining the size of the molecule.  Note that as $b\to 0$, the molecule reverts to a point-particle, so that together with the ``range'' of the potential, $b$ determines the overall distance scale at which the multipole approximation is valid.  For convenience, we set $m_1=m/a$, $m_2=m_3=m$ (where $m$ is a mass scale) so that the center of mass is at $\br=0$ in the body-frame of reference.  Details of the simulation algorithm are provided in other manuscripts.\cite{Patroneinprep,Limecooler,Limecooler2}  

\begin{figure*}
\includegraphics[width=16cm]{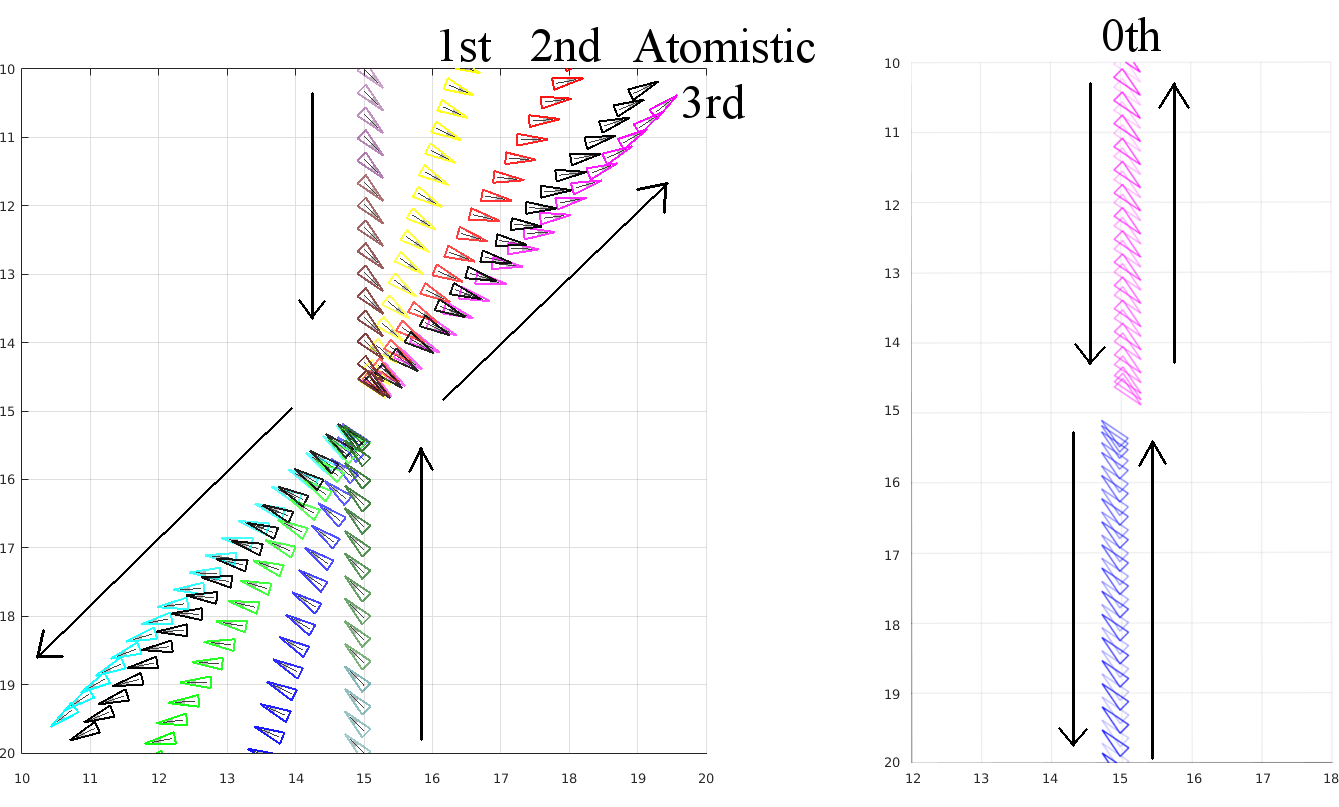}\caption{Top-down view of collision using various approximations to the exact intermolecular potential.  The particle centers-of-mass initially lie along a line parallel to the vertical axis. {\it Left:} Comparison between trajectories computed using the atomistic potential and the 1st, 2nd, and 3rd order approximations.  All trajectories initially overlap and then diverge after the collision. Note, however, that the trajectories converge to the atomistic one with increasing approximation order.  Trajectories in bottom half of figure correspond to the top half when inverted about the line $y=x$ {\it Right:}  Trajectory computed using the zeroth-order approximate potential.  Note the lack of deflection in the horizontal direction.}\label{fig:inplane}
\end{figure*}

Figure \ref{fig:collsion} shows the results of four different simulations of molecules in head-on collisions.  We prescribe initial conditions as follows:
\begin{subequations}
\begin{align}
\bx_1 &= (10,15.05,15.25)^\T \\
\bx_2 &= (20,14.95,14.75)^\T \\
\dot \bx_1 &= (v_1,0,0)^\T \\
\dot \bx_2 &= (-v_1,0,0)^\T \\
\bomega_1 &=\bomega_2=\boldsymbol 0 \\
\bt_1 &= \bt_2 =\boldsymbol {\rm R}_y(\pi/2) \boldsymbol {\rm R}_x(\pi/4) \\
a&=4, \,\,\,b=0.1 \\
m& =0.1 
\end{align}
\end{subequations}
where $v_1>0$ is an adjustable parameter, $\boldsymbol {\rm R}_{\star}(\phi)$ is a rotation about the lab-frame $\star$-axis by an angle $\phi$, and $a$ and $b$ are as defined in Eqs.~\eqref{eq:tr1}--\eqref{eq:tr2}.  We choose $v_1=160$ and set the interatomic potential $u(r)=300 r^{-2}$ to be a simple power law repulsion.
 Visually it is apparent that the zeroth-order potential (which is radially symmetric) does not induce rotation, entirely inconsistent with the physics of the exact potential.  Increasing the approximation order of the potential yields better agreement with the exact trajectory as shown in the bottom two subplots.  Figure \ref{fig:inplane} shows a related collision in which the two particles are in-plane during the collision.  Here $u(r)=45r^{-2}$ and $v_1=55$.  Notably, the zeroth-order potential fails to predict deflection off the $x$-axis, which is inconsistent with the trajectory of the exact potential.  Further analysis of these simulations, including an assessment of the phase-space convergence, will be provided in a manuscript in preparation.\cite{Patroneinprep}

\section{Discussion and Conclusions}

As discussed in the introduction, generalized multipole potentials are useful in that they provide an analytical method for coarse-graining rigid-body systems.  Moreover, a salient feature of the method as it pertains to UQ is that error in the local force-field evaluations can always be estimated by the first omitted term in the expansion.  For example, an approximation to second order in the ratio $\epsilon = |\blk|/\chi \ll 1$ will yield errors that are $\mathcal O(\epsilon^3)$, which may often be neglected for large enough separations.  We anticipate that such asymptotic approximations may be helpful in constructing more formal estimates of uncertainties associated with bulk properties of many-body simulations.

This being said, usefulness of the multipole approximation also relies on the validity of certain  assumptions, which we now discuss.  In particular, its accuracy is only guaranteed under far-field conditions.  In condensed-matter systems we always expect some subset of molecules to violate this condition, e.g.\ nearest-neighbors in close-packed systems.  However, we note that the coarse-grained representation in terms of Eq.~\eqref{eq:cgrep} contains all the atomistic information about the system, albeit in a compressed form.  As a result, it is always possible to use the full potential given by Eq.~\eqref{eq:inversepower} whenever the far-field conditions do not hold. In this way, the generalized multipole potential can be tailored to have prescribed levels of accuracy at different separations, so that the overall potential is accurate to within a desired energy threshold.

A second key assumption of our analysis is its restriction to rigid-body systems.  In general, many of the most useful MD models incorporate bond-lengths that vibrate about some mean distance in order to represent internal kinetic energy of the system.  Such effects are not captured by the potential described in Eq.~\eqref{eq:inversepower}, since the body-frame coordinates are assumed to be constant.  Thus, an extension of our analysis is to add a perturbative correction associated with deviations from these average lengths.  It is likely that such terms will contribute to higher-order corrections in the potential and couple to translational and rotational DOF in non-trivial ways.  Analysis of this scenario is left for future work.

Despite these shortcomings, the generalized multipole potential provides a useful starting point for recasting coarse-grained models as approximations and/or truncations of a fully atomistic model.  The key insight we wish to emphasize is the recognition that {\it a priori} uncertainty quantification becomes straightforward in such circumstances, as the models become fully connected through appropriate limiting processes.  Moreover, such analyses provide benchmarks against which other coarse-graining methods can be understood, given their assumptions in comparison to the multipole potentials.

\bibliography{RDF_rep}
\bibliographystyle{aiaa}
\end{document}